\documentclass[aps,twocolumn,preprintnumbers,superscriptaddress]{revtex4}
\usepackage{graphicx}
\usepackage{amsmath}
\usepackage{amssymb}
\usepackage{color}

\begin{document}

\title{Fermi arc induced vortex structure in Weyl beam shifts}

\author{Udvas Chattopadhyay}

\affiliation{Division of Physics and Applied Physics, School of Physical and Mathematical Sciences,\\
Nanyang Technological University, Singapore 637371, Singapore}

\author{Li-kun Shi}

\affiliation{Institute of High Performance Computing, A*STAR, Singapore 138632,
Singapore}

\author{Baile Zhang}

\affiliation{Division of Physics and Applied Physics, School of Physical and Mathematical Sciences,\\
Nanyang Technological University, Singapore 637371, Singapore}

\affiliation{Centre for Disruptive Photonic Technologies, Nanyang Technological University, Singapore 637371, Singapore}

\author{Justin C.~W.~Song}

\affiliation{Division of Physics and Applied Physics, School of Physical and Mathematical Sciences,\\
Nanyang Technological University, Singapore 637371, Singapore}

\affiliation{Institute of High Performance Computing, A*STAR, Singapore 138632,
Singapore}

\author{Y.~D.~Chong}

\affiliation{Division of Physics and Applied Physics, School of Physical and Mathematical Sciences,\\
Nanyang Technological University, Singapore 637371, Singapore}

\affiliation{Centre for Disruptive Photonic Technologies, Nanyang Technological University, Singapore 637371, Singapore}

\begin{abstract}
  In periodic media, despite the close relationship between geometrical effects in the bulk and topological surface states, the two are typically probed separately. We show that when beams in a Weyl medium reflect off an interface with a gapped medium, the trajectory is influenced by both bulk geometrical effects and the Fermi arc surface states.  The reflected beam experiences a displacement, analogous to the Goos-H\"anchen or Imbert-Fedorov shifts, that forms a half-vortex in the two-dimensional surface momentum space.  The half-vortex is centered where the Fermi arc of the reflecting surface touches the Weyl cone, with the magnitude of the shift scaling as an inverse square root away from the touching-point, and diverging at the touching-point. This striking feature provides a way to use bulk transport to probe the topological characteristics of a Weyl medium.
\end{abstract}

\maketitle

\textit{Introduction.}---One of the most interesting features of wave dynamics in both classical and quantum media is that wavepacket trajectories are not determined solely by the dispersion relation, but can also be influenced by the internal structure of the underlying wavefunctions.  For instance, the equations of motion of a wavepacket in a periodic medium can include an ``anomalous velocity'' term tied to the Berry connection of the Bloch functions \cite{KarplusLuttinger1954, Chang1995, Chang1996, Nagaosa2004, Yidong2010, Niu2010}.  Another example, originating from the field of optics, involves the displacement of a light beam reflecting off a surface \cite{Bliokh2013}.  A lateral displacement is called a Goos-H\"anchen (GH) shift \cite{GH_2}, while a transverse one is called an Imbert-Fedorov (IF) shift \cite{Imbert}, and both originate from the polarization degree of freedom of electromagnetic waves
\cite{Player1987, Fedoseyev1998}. In electronic systems such as strained graphene, similar shifts can be induced by pseudospin degrees of freedom, and are predicted to have observable effects on transport in heterojunction devices \cite{Beenakker2009, Chen2013}.

Another phenomenon intimately linked to the internal structure of wavefunctions is the existence of topological surface states, which arise from subtle windings of the Bloch functions in momentum space \cite{Hatsugai1993}.  For example, three-dimensional Weyl media feature linear band-crossing points called Weyl points that act as monopole sources of Berry curvature in momentum space~\cite{TurnerVishwanath2011}, and cause several nontrivial bulk dynamical effects \cite{Burkov2012, Son2013, Burkov2014, Parameswaran2014, Yang2015, Jiang2015, Phuan2015, Huang2015, Jiang2016, Wang2017}. The net Berry flux between pairs of Weyl points guarantees the existence of ``Fermi arcs'' of surface states along any interface with a gapped medium~\cite{TurnerVishwanath2011}. Weyl media have been realized in multiple venues including photonic crystals and waveguide arrays \cite{Joannopoulos2015WeylObserv, CTChan2016WeylObs, Rechtsman2017WeylObs}, materials such as TaAs \cite{Lv2015TaAs, ZahidFermiArc2015, Xu2016TaP, ArmitageRMP2018}, as well as acoustic, mechanical, and electric metamaterials \cite{WeylObservPhononic, CTChan2015Acoustic, Chen2018Acoustic, Wang2016, Rocklin2016, Lee2018}.

Despite this bulk-edge correspondence, the nontrivial bulk dynamics of Weyl media (a band geometric property) and Fermi arc surface states (a topological property) have largely been probed separately. The former has been studied using spectroscopic tools like angle-resolved photoemission~\cite{ZahidFermiArc2015}, whereas the latter has been studied through bulk transport effects such as negative magnetoresistance \cite{Son2013,Phuan2015,Huang2015} and anomalous Hall conductivity \cite{Burkov2014}.  Another interesting bulk phenomenon in Weyl media involves the GH and IF shifts experienced by a wavepacket undergoing partial reflection off a potential step in the bulk~\cite{Yang2015, Jiang2015, Jiang2016, Wang2017}. These shifts have been attributed to the Weyl medium's spinor degree of freedom, and the direction of the IF shift has been shown to be determined by the sign of the Weyl point's Berry flux.  However, no direct connection to the Fermi arc surface states was identified.

In this Letter, we show that when a beam in a Weyl medium reflects off an interface with a gapped medium, the displacement of the reflected beam exhibits an anomalous half-vortex structure in momentum space.  This effect provides a bulk probe of the topological Fermi arc at the reflecting surface.  Previous studies into GH and IF shifts in Weyl media dealt with \textit{partial} reflection off a potential step separating two Weyl media \cite{Yang2015, Jiang2015, Wang2017}.  In contrast, we consider \textit{total} reflection off an interface with a medium that is gapped (i.e., supporting no propagating waves at the operating energy).  Unlike a potential step, such an interface features a Fermi arc extending outward from the Weyl cone in the two-dimensional surface momentum space \cite{TurnerVishwanath2011}.  The real-space displacement of the reflected beam, $\vec{\Delta}$, varies with the direction of the incident beam, which is characterized by its average in-plane momentum $\vec{K}_\perp$.  We show that $\vec{\Delta}(\vec{K}_\perp)$ circulates around $\vec{K}_{\mathrm{fa}}$, the point on the boundary of the Weyl cone touched by the Fermi arc, and its magnitude scales as an inverse square root, $|\vec{\Delta}| \sim |\vec{K}_\perp - \vec{K}_\mathrm{fa}|^{-1/2}$.  This behavior is observed in two different Hamiltonian models with distinct boundary condition implementations, indicating that it is generic.  In microwave photonic crystals \cite{Joannopoulos2015WeylObserv}, the magnitude of the predicted shift is several multiples of the lattice constant under realistic conditions.  The displacement accumulates over successive reflections off two parallel surfaces, and thus affects the effective velocity of propagation \cite{Beenakker2009} in films of Weyl medium.  To our knowledge, this is the first prediction of the Fermi arc having observable effects on beam trajectories in Weyl media, which may inspire further studies of the physical effects of Fermi arcs.

\begin{figure}
  {\centering \includegraphics[width=0.45\textwidth]{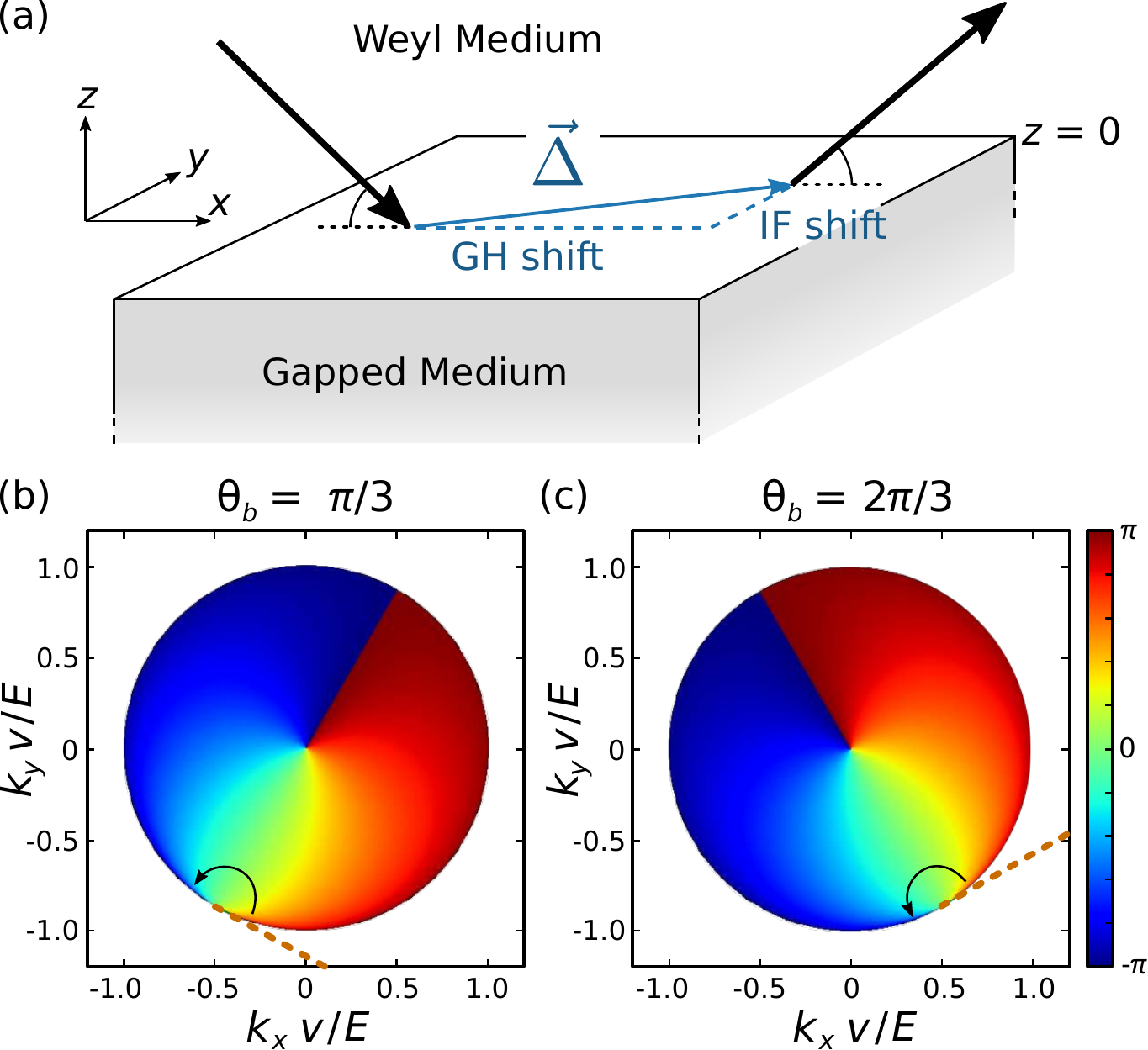}}
  \caption{(a) Schematic of the reflection setup.  A beam is incident
    from a Weyl medium in the space $z > 0$, and reflects off a gapped
    medium at $z < 0$.  The vector $\vec{\Delta}$ denotes the total
    displacement of the reflected beam.  (b)--(c) Map of the
    reflection phase $\phi$ experienced by an incident plane wave,
    versus the in-plane wavenumbers $k_x$ and $k_y$ (normalized to
    $E/v$).  Results are shown for two different values of $\theta_b$,
    which governs the boundary condition.  The Fermi arc is denoted by
    dashes, and $\phi$ exhibits a $-2\pi$ phase shift during a
    half-encirclement of the Fermi arc's touching-point (black
    arrow). }
  \label{fig:setup}
\end{figure}

\textit{Plane waves in a Weyl medium.}---We consider the setup shown in Fig.~\ref{fig:setup}(a), where a Weyl medium occupies the space $z > 0$ with a gapped medium in $z < 0$.  A monochromatic beam of energy $E$ is incident from the Weyl medium, and reflects off the $z = 0$ surface.  The reflected beam can experience a displacement, denoted by a vector $\vec{\Delta}$ parallel to the $x$-$y$ plane.  The reflection is total, for $E$ lies in the gap of the $z < 0$ medium.

The eigenmodes of the bulk Weyl medium are described by a $2\times 2$ Hamiltonian $H_w= \sum_j v_j k_j \sigma_j$, where for each direction $j \in \{1,2,3\}$, $v_j$ is the phase velocity, $k_j$ is the wavenumber, and $\sigma_j$ is a Pauli matrix.  (In quantum mechanical contexts, we set $\hbar = 1$.)  The Weyl point possesses a chirality invariant $C = \textrm{sgn}(v_xv_yv_z)$, which can only be altered by annihilation with another Weyl point \cite{TurnerVishwanath2011}.  We henceforth take $v_x=v_y=v_z=v$, so that $C= \textrm{sgn} (v)$.  For given $\vec{k}$, the modal eigenenergy (or eigenfrequency) is $E = v|\vec{k}|$, and the wavefunction is a superposition of two basis wavefunctions with coefficients given by the spinor components of the envelope function $\psi(\vec{k},\vec{r}) = \Psi(\vec{k}) \, \exp\big(i \vec{k}\cdot\vec{r}\big)$, where $\Psi(\vec{k})$ is an eigenvector of $H_w(\vec{k})$ and $\vec{r} \equiv (x,y,z)$.

For incident and reflected plane waves (not beams), the eigenvectors are taken to be
\begin{equation}
  \Psi_i = \frac{1}{\sqrt{1+\eta_-^2}}
  \begin{bmatrix}
    1\\
    \eta_- e^{i\alpha}
  \end{bmatrix}, \;\;
  \Psi_r = \frac{e^{i\phi}}{\sqrt{1+\eta_+^2}}
  \begin{bmatrix}
    1\\
    \eta_+ e^{i\alpha}
  \end{bmatrix},
  \label{eigir}
\end{equation}
where
\begin{align}
  \begin{aligned}
  \alpha &= \tan^{-1}\left(\frac{k_y}{k_x}\right), \;\;
  \eta_{\pm} = \sqrt{\frac{E-vk_z^{\pm}}{E+vk_z^{\pm}}}, \\
  k_z^{\pm} &= \pm \sqrt{(E/v)^2 - |\vec{k}_\perp|^2}.
  \label{eq:alpha}
  \end{aligned}
\end{align}
The $k_z < 0$ ($k_z > 0$) branch is chosen for the incident (reflected) wave, and $\vec{k}_\perp$ is the projection of $\vec{k}$ onto the $k_x$-$k_y$ plane.  The $\exp(i\phi)$ factor is a reflection coefficient, determined by the boundary condition at $z = 0$.

From the Hermiticity of the real-space Hamiltonian, one can show \cite{EdwardBCWeyl,AkhmerovBCDirac,BCWeyl} that the boundary of a Weyl medium is characterizable by a single real angular parameter $\theta_b \in [0, 2\pi]$, such that
\begin{equation}
  \begin{bmatrix} 1 & e^{-i\theta_b(\vec{k}_\perp)} \end{bmatrix}
  \psi_{\mathrm{tot}}\Big|_{z=0} = 0,
  \label{BCe}
\end{equation}
where $\psi_{\mathrm{tot}}$ is the sum of the incident and reflected envelope functions.  Note that $\theta_b$ may vary with $\vec{k}_\perp$.  Hence,
\begin{equation}
  e^{i\phi} = -\sqrt{\frac{1+\eta_+^2}{1+\eta_-^2}} \;
  \frac{1+\eta_- e^{i\alpha} e^{-i\theta_b}}{1+\eta_+ e^{i\alpha} e^{-i\theta_b}}.
\end{equation}
Eq.~\eqref{BCe} also yields surface states, which have the form $\psi_{s} \propto e^{-\kappa z} [e^{-i\theta_b}, -1]^T$ for $z > 0$, where $\kappa = k_x \sin\theta_b-k_y \cos\theta_b \ge 0$.  These lie along a Fermi arc \cite{TurnerVishwanath2011} given by
\begin{equation}
  E = -k_x \cos\theta_b-k_y \sin\theta_b.
  \label{farc}
\end{equation}
In this model, the Fermi arc extends to infinity, since there is only a single Weyl cone.

The resulting reflection phase is shown in Fig.~\ref{fig:setup}(b)--(c), for two representative cases where $\theta_b$ is a constant independent of $\vec{k}_\perp$.  The color map gives the values of $\phi$ within the circular domain $|\vec{k}_\perp| \le E/v \equiv K_W$, which is a section of the Weyl cone.  The Fermi arc lies outside the cone and touches its boundary tangentially.  The touching-point, and the orientation of the Fermi arc, depend on the choice of $\theta_b$.  We see that $\phi$ winds by $2\pi$ during a half-encirclement around the Fermi arc touching-point.

These features can be understood by considering the Weyl cone section's boundary, $|\vec{k}_\perp| = K_W$, which is parameterized by the polar angle $\alpha$ [Eq.~\eqref{eq:alpha}].  As we approach the boundary from the inside (i.e., real $k_z \rightarrow 0$), the expressions for $\Psi_i$ and $\Psi_r$ in Eq.~\eqref{eigir} become linearly dependent, so that $\psi_{\mathrm{tot}} \propto (1+e^{i\phi})\Psi_i$ and $\eta_\pm \rightarrow 1$.  Hence, the boundary condition \eqref{BCe} can be satisfied in two ways: (i) $\phi = \pi$, so that $\psi_{\mathrm{tot}}$ vanishes, or (ii) $\alpha - \theta_b(\vec{k}_\perp) = \pi \;(\textrm{mod} \;2\pi)$, so that the vectors in Eq.~\eqref{BCe} are orthogonal.  Case (ii) corresponds to the touching-point of the Fermi arc, and when it is satisfied $\phi$ is undefined rather than being equal to $\pi$ as in case (i).  Evidently, this occurs at a minimum of $|m-1|$ distinct values of $\alpha$, where $m$ is the winding number of the $\theta_b(\vec{k}_\perp)$ function.  We deduce that $m = 0$, which includes the case where $\theta_b$ is a constant, describes the class of boundaries between a Weyl medium and a trivial gapped medium.  Then there is a minimum of one touching-point, and by considering the points just inside the boundary we find that $\phi$ winds by $2\pi$ during a half-encirclement of the touching-point \cite{SM}.

\textit{Beams in Weyl media.}---A monochromatic beam of energy $E$ can be described as a superposition of planar eigenmodes, with wavenumbers $\vec{k}$ constrained by $E = v|\vec{k}|$.  The envelope function has the form
\begin{equation}
  \Psi(\vec{r}, \vec{K}_\perp)
  = \int d^2k_\perp \, g(k_x-K_x) \,g(k_y-K_y)\, \psi(\vec{k},\vec{r}),
  \label{beam}
\end{equation}
where $\vec{k}_\perp \equiv (k_x, k_y)$ denotes the in-plane wavevector of each eigenmode, $\vec{K}_\perp$ is the central value for the beam's in-plane wavevector, $g(k_j - K_j)$ is a $k$-space envelope function, and $\psi(\vec{k},\vec{r})$ describes a planar eigenmode.  We let each $g$ be a Gaussian function of unit area, zero mean, and standard deviation $\sigma_k$.  For each $E$, the choice of sign for $k_z$ is determined by the beam direction.

We plug the incident and reflected eigenmodes [Eq.~\eqref{eigir}] into Eq.~\eqref{beam}, and expand $\phi$ and $\alpha$ to lowest order, so as to compare the centers of the incident and reflected beams in the $z=0$ plane.  This results in the following formula for the displacement of the reflected beam:
\begin{equation}
  \vec{\Delta}(\vec{K}_\perp)
  = \left[-\nabla_{k_\perp} \phi \, + \,
    \frac{\eta_{-}^2-1}{\eta_{-}^2+1} \, \nabla_{k_\perp} \alpha
    \right]_{\vec{k}_\perp = \vec{K}_\perp}.
  \label{shift}
\end{equation}
Here, $\nabla_{k_\perp}$ denotes the in-plane $k$-space derivative, and $\vec{K}_\perp$ is the mean wave-vector for the incident beam.  Although $\phi$, $\eta_\pm$, and $\alpha$ are based on an eigenmode gauge choice [Eq.~\eqref{eigir}], the shift \eqref{shift} is gauge-independent.  For details, see the Supplemental Material \cite{SM}.

\begin{figure} \centering
  \includegraphics[width=0.49\textwidth]{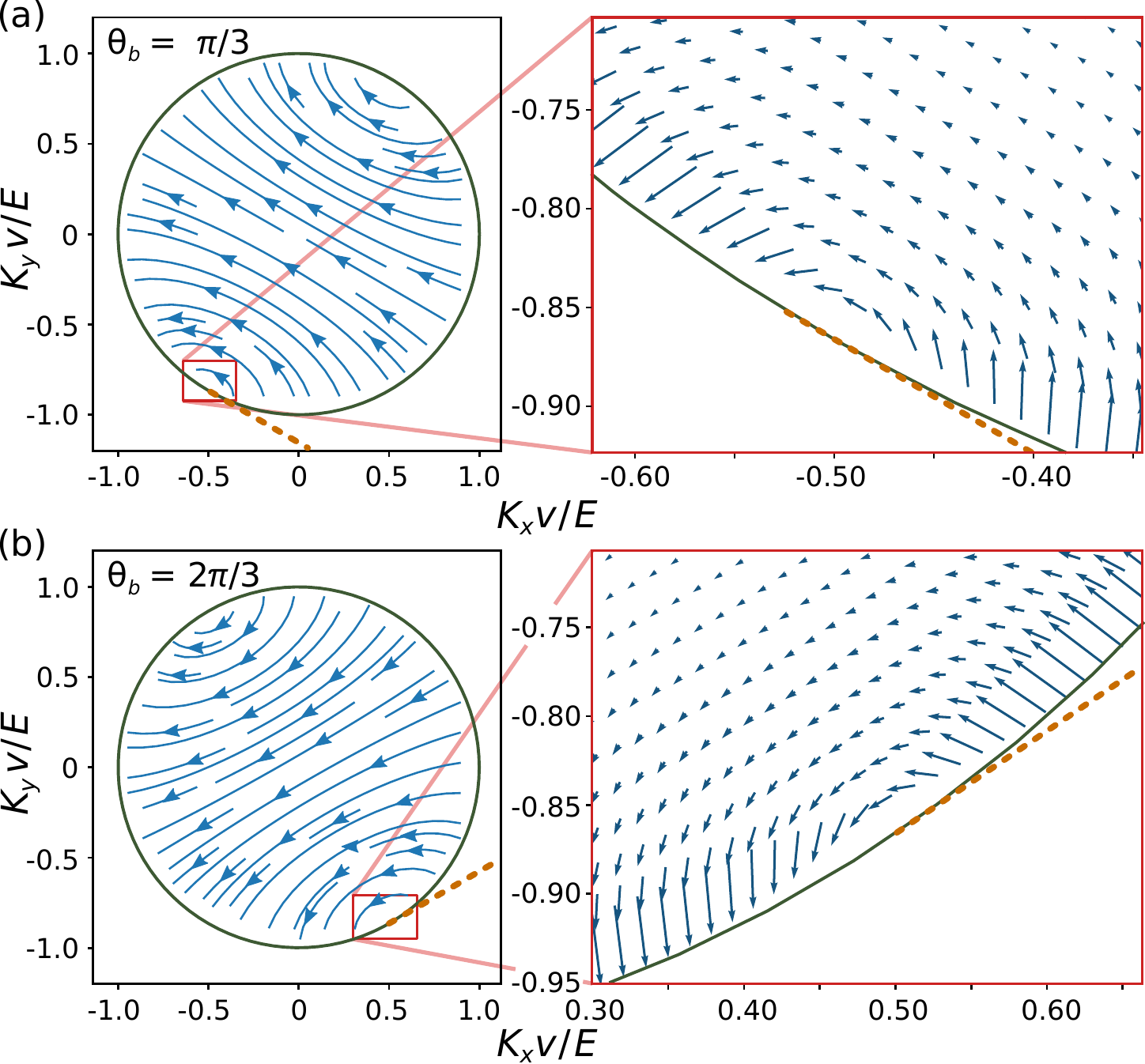}
  \caption{Maps of the beam shift $\vec\Delta$ versus the mean
    in-plane wavenumbers $K_x$ and $K_y$ (normalized to $E/v$) for the
    incident beam.  Results are shown for two values of the boundary
    parameter: (a) $\theta_b = \pi/3$ and (b) $\theta_b = 2\pi/3$.  In
    each case, the left panel shows a streamline plot indicating the
    direction of $\vec\Delta$ but not its magnitude; the right panel
    shows a quiver plot indicating both the direction and magnitude of
    $\vec\Delta$, over a region of $K$-space surrounding the Fermi arc
    touching-point.  The dashes indicate the Fermi arc.}
  \label{Shift_plot}
\end{figure}

Fig.~\ref{Shift_plot} shows streamline and quiver plots of $\vec{\Delta}$ as a function of $\vec{K}_\perp$. We find that the direction of $\vec{\Delta}$ winds by $\pi$ during a half-encirclement of the Fermi arc touching-point $\vec{K}_{\mathrm{fa}}$, and moreover that its magnitude scales as
\begin{equation}
  |\vec{\Delta}| \sim \left|\vec{K}_\perp - \vec{K}_{\mathrm{fa}} \right|^{\,-1/2}.
  \label{Vorticity}
\end{equation}
By analogy with the concept of free vortices from fluid mechanics, we say that $\vec{\Delta}$ exhibits a ``half-vortex'' structure centered at $\vec{K}_{\mathrm{fa}}$.  Note that the shift direction is not purely lateral (GH-like) or transverse (IF-like), but depends on the orientation of the incident beam relative to the Fermi arc.  We interpret the formal divergence of $|\vec{\Delta}|$ at $\vec{K}_{\mathrm{fa}}$ to mean that the presence of the Fermi arc causes the beam to deviate significantly from an ideal ray-like trajectory.  The scaling \eqref{Vorticity} is observed numerically for different directions stretching away from $\vec{K}_{\mathrm{fa}}$, and we can show that it comes from the $-\nabla_{k_\perp} \phi$ term in Eq.~\eqref{shift}; details are given in the Supplemental Material \cite{SM}.  The analysis also shows that the $\phi$ vortex at the $k$-space origin [Fig.~\ref{fig:setup}(b)--(c)], a feature noted by several previous authors \cite{Hailong2016, Gao2016, CTChan2017}, does \textit{not} yield a vortex in $\vec\Delta$, as the two terms in Eq.~\eqref{shift} have similar magnitudes and opposite vorticities and cancel.  Near the Weyl cone, however, the $\nabla_{k_\perp} \alpha$ term is negligible and only $\nabla_{k_\perp} \phi$ contributes.

The streamline plots in Fig.~\ref{Shift_plot} also show that $\vec{\Delta}$ winds around a point $\vec{K}_{\mathrm{opp}}$ \textit{opposite} to the Fermi arc touching-point.  However, around this point the magnitude of the beam shift scales as $|\vec{\Delta}| \sim |\vec{K}-\vec{K}_{\mathrm{opp}}|^{1/2}$.  The vanishing of $\vec{\Delta}$ at $\vec{K}_{\mathrm{opp}}$ means that deviations from ideal beam trajectories, due to the internal spinor degrees of freedom, become negligible in this region of momentum space.

\begin{figure*}
  \centering
  \includegraphics[width=0.8\textwidth]{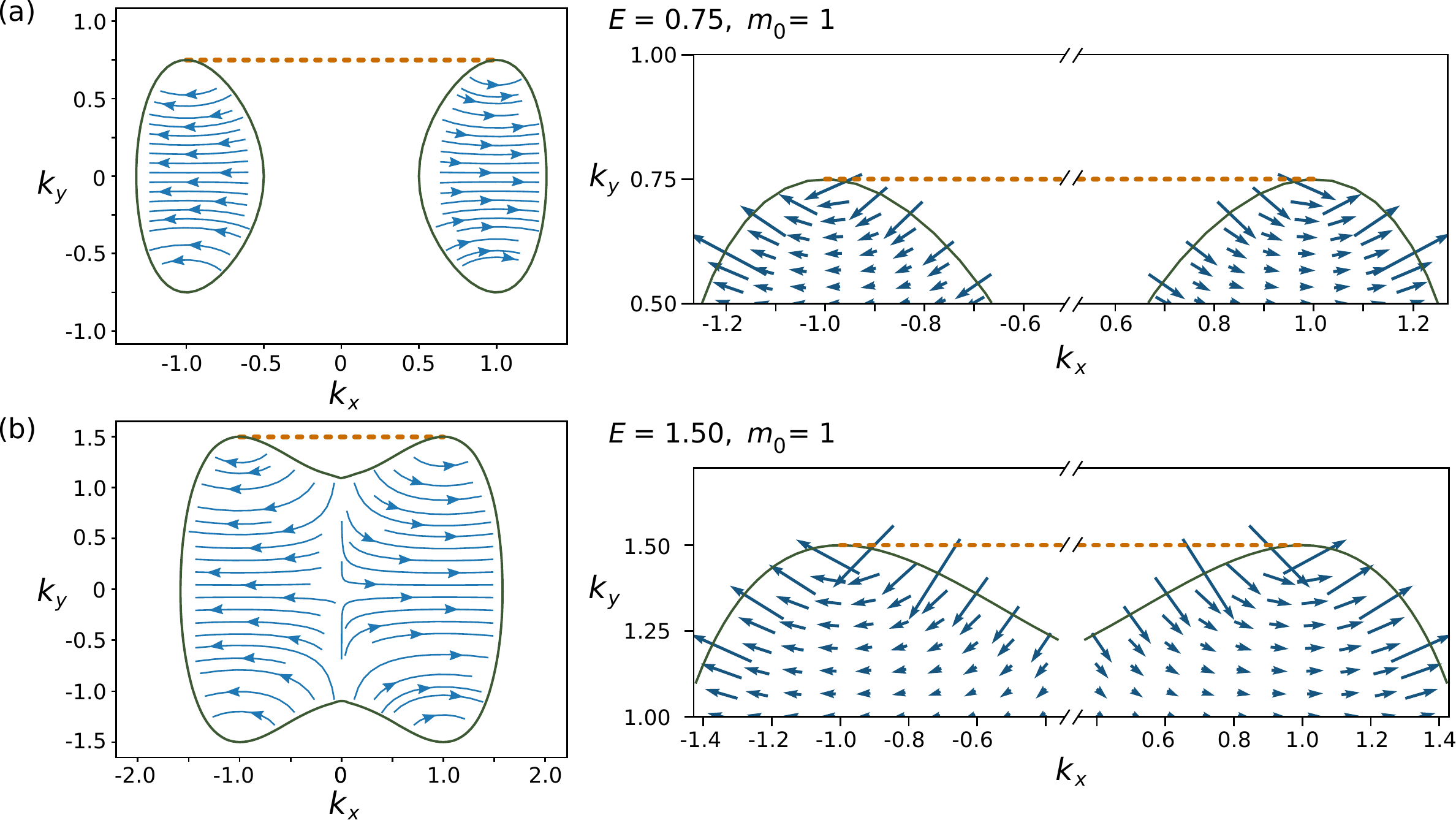}
  \caption{Maps of the beam shift $\vec\Delta$, for the quadratic
    Hamiltonian \eqref{quadHamiltonian} with $v = 1$, $m_0 = 1$ and
    $m_1 = 100$.  The horizontal and vertical axes
    are the mean in-plane wavenumbers $K_x$ and $K_y$ (normalized to
    $E/v$) for the incident beam.  Results are shown for (a) $E =
    0.75$, where the Hamiltonian describes a pair of Weyl cones, and
    (b) $E = 1.5$, where the two Weyl cones have merged into a single
    band.  The left panels show a streamline plot of $\vec{\Delta}$,
    while the right panels show a quiver plot of $\vec{\Delta}$ near
    the Fermi arc touching-points.  The dashes indicate the Fermi
    arc.}
  \label{Quad_shift_plot}
\end{figure*}

\textit{Paired Weyl cones.}---To show that the above results are not model-specific, we consider an alternative model described by the quadratic Hamiltonian \cite{QuadWeylFermi, GorbarFermiArc}
\begin{equation}
  H = v\begin{bmatrix}
  -k_y & \beta(k_x^2-m) - i  k_z  \\
  \beta(k_x^2-m) + ik_z & k_y
  \end{bmatrix}.
  \label{quadHamiltonian}
\end{equation}
This has dispersion $E = \pm v\sqrt{\beta^2 (k_x^2-m)^2 + k_y^2 + k_z^2}$, and exhibits either paired Weyl points or a complete bandgap, depending on the choice of $m$.  In the region $z > 0$, we set $m = m_0 > 0$, so that there are Weyl points at $\vec{k} = [\pm\sqrt{m_0}, \,0, \, 0]$; there are two Weyl cones for small $|E|$, which merge into a single band for larger $|E|$.  For $z < 0$, we let $m = -m_1 < 0$ to ensure a complete gap for $|E| < v\beta|m_1|$, and set $E$ within this gap.

The calculation of the beam shift proceeds along similar lines.  The form of the planar eigenmodes in the $z > 0$ region is similar to Eq.~\eqref{eigir}, and evanescent for $z < 0$.  Instead of using the boundary equation \eqref{BCe}, we require the components of $\psi_{\mathrm{tot}}$ to be continuous at $z = 0$.  The Fermi arc is then found to be the line segment $|k_x| < \sqrt{m_0}$, $k_y = E/v$.  Finally, we construct beams similar to \eqref{beam} and calculate the reflected beam displacement $\vec{\Delta}$.

The resulting plots of $\vec{\Delta}$ versus $\vec{K}_\perp$ are shown in Fig.~\ref{Quad_shift_plot}.  The behavior is highly similar to the single-cone case, despite key differences in the calculation---not only in the Hamiltonian, but also the boundary condition implementation.  There are now two Weyl cones; the cone boundaries are no longer circular, but nonetheless each cone has its own Fermi arc touching-point, and $\vec{\Delta}$ exhibits a half-vortex structure around each touching-point.  The winding direction is opposite in the two Weyl cones, consistent with their opposite chiralities, and a numerical fit shows that the $|\vec{\Delta}| \sim |\vec{K}_\perp - \vec{K}_\mathrm{fa}|^{-1/2}$ scaling holds near the touching-points.  The behavior persists even at large values of $E$ where the cones merge into a single band.  These results indicate that these features of the beam shift are generic to Weyl media, and are not qualitatively altered by model-specific bandstructure features.

\textit{Discussion.}---The easiest way to observe the predicted beam shift may be to use a classical Weyl medium, such as a microwave-scale photonic crystal of the sort implemented by Lu \textit{et al.}~\cite{Joannopoulos2015WeylObserv}.  In a microwave experiment, a phase array can be used to generate the incident beam, and a metal surface can serve as a reflector \cite{Pozar}.  In Ref.~\cite{Joannopoulos2015WeylObserv}, the lattice constant is $13.4\,\textrm{mm}$ and the Weyl points occur at frequency $f \approx 11.3\,\textrm{GHz}$, with $v \approx 7\times10^7\,\textrm{ms}^{-1}$.  Operating 5\% above the Weyl point frequency ($\delta f \approx 0.57\,\textrm{GHz}$), the Weyl cone section has radius $K_W \approx 50\,\textrm{m}^{-1}$.  For a beam of momentum-space width $\sigma_k \approx 5\,\textrm{m}^{-1}$ (real-space width $\approx 100\,\textrm{mm}$), with incident beam direction such that $|\vec{K}_\perp - \vec{K}_{\textrm{fa}}| \approx 15\,\textrm{m}^{-1}$, the envelope has negligible overlap with the boundary of the Weyl cone.  The model of Eqs.~\eqref{eigir}--\eqref{shift} then predicts $|\vec{\Delta}| \sim 46\,\textrm{mm}$, which should be easily observable.

In solid state systems, these GH- and IF-like shifts may provide a probe for topological Fermi arc effects in Weyl semimetals and related materials.  In a Weyl semimetal thin film bounded above and below by insulating media, a beam or traveling wavepacket undergoes repeated reflections off the two parallel surfaces, and a straightforward calculation shows that the shifts accumulate rather than canceling out~\cite{SM}, producing an anomalous boost to the in-plane motion.  GH shifts have been predicted to contribute to two-terminal conductance in graphene {\it p-n} interfaces~\cite{Beenakker2009}, and likewise Fermi arc-induced shifts may be detectable via transport properties in thin films of Weyl semimetals.  Electrons in topological semimetals have been shown to have very high mobilities \cite{Huang2015,Shekhar2015}, allowing multiple reflections to fit within a transport mean free path.  Finally, although repeated reflections might change the beam profile, owing to higher-order terms neglected in the above calculations, one can show that the beam shift formula \eqref{shift} remains valid even if the beam profile is distorted, as long as it remains a real function with a single peak in real space \cite{SM}.

\textit{Acknowledgments}---This work was supported by the Singapore MOE Academic Research Fund Tier 2 Grant MOE2015-T2-2-008, and the Singapore MOE Academic Research Fund Tier 3 Grant MOE2016-T3-1-006. J.C.W.S acknowledges the support of the Singapore National Research Foundation (NRF) under NRF fellowship award NRF-NRFF2016-05.

\bibliography{references}

\pagebreak

\begin{widetext}

\renewcommand{\thefigure}{S\arabic{figure}}
\renewcommand{\theequation}{S\arabic{equation}}
\setcounter{figure}{0}
\setcounter{equation}{0}

\begin{center}
{\Large Supplemental Material for \\Fermi arc induced vortex structure in Weyl beam shifts}    
\end{center}

In this supplement, we discuss the winding behavior of the planar reflection coefficient $\phi$; derive analytical formulas for the shift of Gaussian beams in a Weyl medium, and compare them to numerical results; derive the eigenvectors, reflection coefficients, and beam shifts in the quadratic Hamiltonian model; and derive the accumulation of beam shifts in a thin film geometry.  Unless otherwise specified, it is assumed that $v=1$ and that $\theta_b$ is a constant.

\subsection{Reflection phase winding around Fermi arc touching-point}

\begin{figure}[b] \centering 
\includegraphics[width=0.45\textwidth]{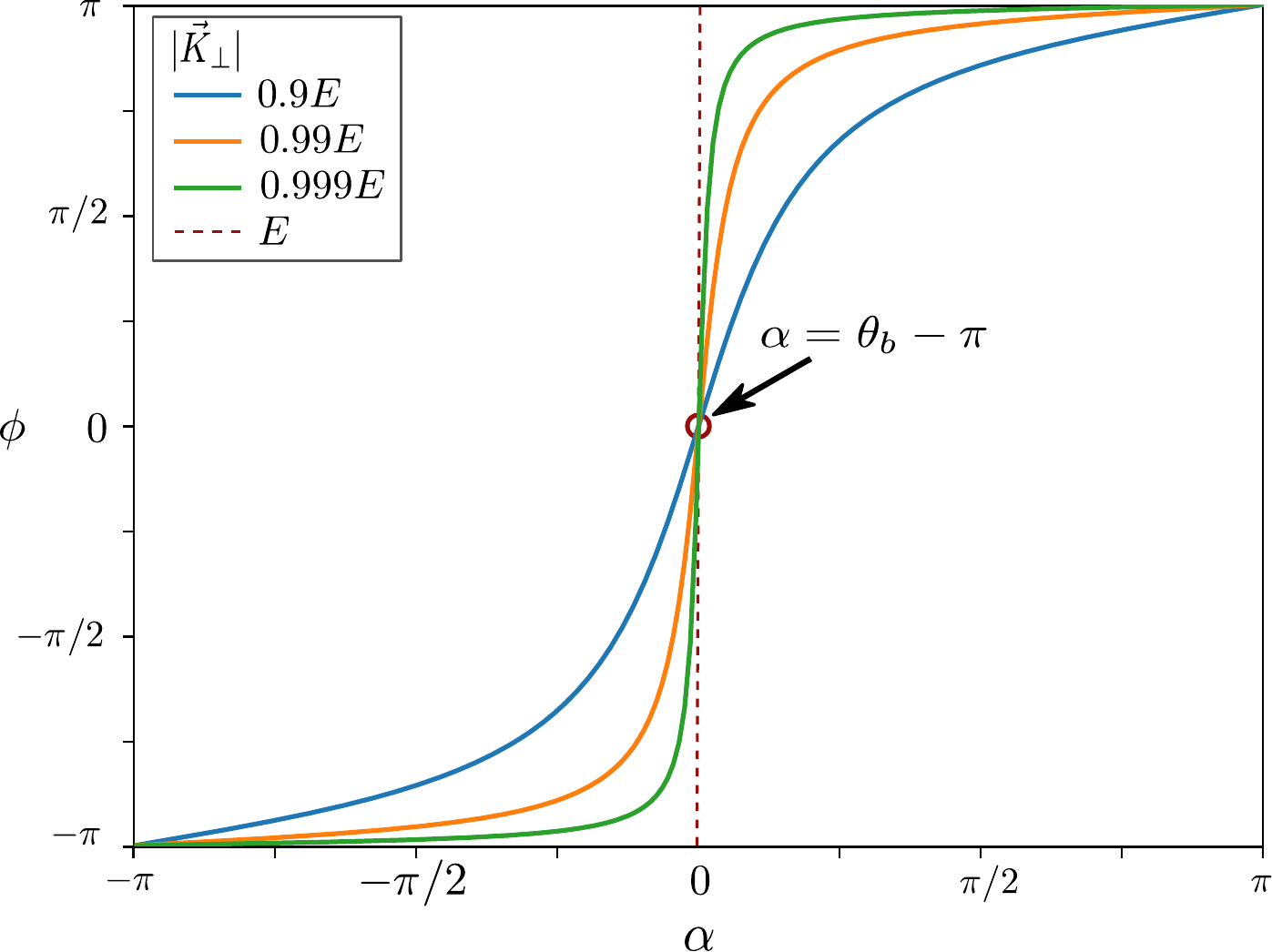}
\caption{Reflection phase $\phi$ versus $\alpha$ (the polar angle in the two-dimensional $k$-space of the incident beam), for $E=1$, $\theta_b = \pi$, and different values of $|\vec{K}_\perp|$ close to the Weyl cone boundary.  The Fermi arc touching-point occurs at $\alpha = 0$.  Exactly at the boundary ($|\vec{K}_\perp| = E$), Eq.~\eqref{bc} states that $\phi = \pm \pi$ for all $\alpha \ne 0$ and is undefined at $\alpha = 0$ (being a step function).}
\label{Winding}
\end{figure}

As mentioned in the main text, the non-trivial behaviour of the reflection phase near the Fermi arc touching-point can be understood by considering the limiting case $k_z^{\pm} \rightarrow 0$. For $k_z^{\pm}=0$, the incident and reflected waves are linearly dependent and the total (envelope) wavefunction is
\begin{equation}
\psi_{\mathrm{tot}} = \frac{1}{\sqrt{2}}(1+e^{i\phi}) 
\begin{bmatrix}
  1 \\
  e^{i\alpha}
\end{bmatrix}.
\end{equation}  
The boundary condition (Eq.~3 of main text) then implies
\begin{equation}
  (1+e^{i\phi})(1+e^{i(\alpha-\theta_b)}) = 0,
  \label{bc}
\end{equation}
which can be satisfied by (i) $\phi=\pi$ and (ii) $\alpha-\theta_b =\pi \;(\textrm{mod} \;2\pi)$. Fig. \ref{Winding} shows $\phi$ as a function of $\alpha$ parametrizing circular patches of different radius close to the boundary. As the boundary $|\vec{K}_\perp|=E$ is approached, $\phi$ becomes ill-defined and $\alpha=\theta_b -\pi$ defines the Fermi arc touching-point. This is consistent with the behavior shown in Fig.~1(b)--(c) of the main text.

To analytically derive the direction in which $\phi$ winds, consider momenta close to the boundary, such that $k_z^\pm = \pm \delta k_z$ where $\delta k_z$ is a small positive number. Let $q = \delta k_z/E$, and expand $\eta_-$ up to linear order in $q$:
\begin{equation}
\eta_- = \sqrt{\frac{1+q}{1-q}} \approx 1+q, \qquad \eta_-^2 \approx 1+2q.
\end{equation}
Then the reflection coefficient is
\begin{align}
  \begin{aligned}
    e^{i\phi} &= -\frac{1+\eta_- e^{i\beta}}{\eta_- + e^{i\beta}} \\
    &= -1 - \frac{2iq\sin \beta}{1+\cos \beta} + O(q^2),
  \end{aligned}
\label{Ref_phase_approx}
\end{align}
where $\beta=\alpha-\theta_b$.  Hence, to leading order,
\begin{equation}
  \sin \phi \approx -\frac{2q\sin \beta}{1+ \cos \beta}.
  \label{img}
\end{equation}
At exactly $\beta = \pm \pi$, the denominator diverges and the approximation breaks down.  For small angle deviations, $\beta=\pi+\delta\beta$, we find that $\sin \phi \approx 2q\sin (\delta \beta)$, so that $\delta \phi$ switches sign with $\delta \beta$.  Note that $q > 0$ for the upper cone, and $q < 0$ for the lower cone, so the two cones have opposite windings.

\subsection{Simple derivation of gaussian beam shifts}

Let $f$ be a function of the in-plane momenta $\vec{k}_\perp = (k_x,k_y)$.  We take a gaussian beam of the form 
\begin{equation}
\Psi = \frac{1}{2\pi\Delta_x\Delta_y}\int_{-\infty}^{\infty} dk_xdk_y \; e^{-\frac{(k_x-K_x)^2}{2\Delta_x^2}} e^{-\frac{(k_y-K_y)^2}{2\Delta_y^2}} e^{i f(x,y,k_x,k_y)} e^{i k_x x + ik_y y}.
\label{Gauss_int}
\end{equation}
To find the center of this beam, expand $f(x, y, k_x,k_y)$ about the mean wave-vector $\vec{K}_\perp=(K_x,K_y)$ as
\begin{equation}
f(x,y,k_x,k_y) \approx f(x,y,K_x,K_y) + \frac{\partial f}{\partial k_x}\Big|_{\vec{K}_\perp} (k_x-K_x)+ \frac{\partial f}{\partial k_y}\Big|_{\vec{K}_\perp} (k_y-K_y).
\label{Taylor_expansion}
\end{equation}
Here, only terms up to first order are retained. The integral \eqref{Gauss_int} can be written as
\begin{equation}
\Psi = C \int_{-\infty}^{\infty} dk_x \;\xi(k_x)e^{ik_xx}
\int_{-\infty}^{\infty} dk_y\; \xi(k_y) e^{ik_yy}
\end{equation}
where 
\begin{equation}
C=\frac{\exp[if(x,y,K_x,K_y)]}{2\pi\Delta_x\Delta_y} , \quad
\xi(k_s) = \exp(-\frac{(k_s-K_s)^2}{2\Delta_s^2}) \exp(i \frac{\partial f}{\partial k_s}\Big|_{\vec{K}_\perp} (k_s-K_s)).
\end{equation}
Evaluating the integrals, we obtain
\begin{equation}
\Psi = \exp\Big(if+iK_xx+iK_yy\Big) \exp\left[-\frac{\Delta_x^2}{2}\left(x+\frac{\partial f}{\partial k_x}\right)^2\right] \exp\left[-\frac{\Delta_y^2}{2}\left(x+\frac{\partial f}{\partial k_y}\right)^2\right].
\end{equation}
We find that the probability amplitude $|\Psi|^2$ is a gaussian in real space, centered at
\begin{equation}
  \vec{R} = \left[-\frac{\partial f}{\partial k_x}, 
    -\frac{\partial f}{\partial k_y} \right]_{\vec{k}_\perp = \vec{K}_\perp}.
\end{equation}

For the particular case of Eq.~(1) of the main text, the two components of the incident beam are centered at $(0,0)$ and $\big(-\partial_{k_x} \alpha, -\partial_{k_y} \alpha\big)_{\vec{k}_\perp = \vec{K}_\perp}$ and the center of the beam is calculated by taking weighted averages over the pre-factors:
\begin{equation}
  \vec{R}_{i} = \left(-\frac{\eta_-^2}{1+\eta_-^2} \nabla_{\vec{k}_\perp} \alpha \; 
  \right)_{\vec{k}_\perp = \vec{K}_\perp},
\end{equation}
where, as in the main text,
\begin{align}
  \alpha &= \tan^{-1}\left(\frac{k_y}{k_x}\right), \;\;
  \eta_{\pm} = \sqrt{\frac{E-vk_z^{\pm}}{E+vk_z^{\pm}}}, \\
  k_z^{\pm} &= \pm \sqrt{(E/v)^2 - |\vec{k}_\perp|^2}.
\end{align}
Similarly, the reflected beam is centered at
\begin{equation}
  \vec{R}_{r} = \left(
  -\frac{1}{1+\eta_+^2}\nabla_{\vec{k}_\perp}\phi
  -\frac{\eta_+^2}{1+\eta_+^2}
  \nabla_{\vec{k}_\perp} (\alpha+\phi) \right)_{\vec{k}_\perp = \vec{K}_\perp}.
\end{equation}
The shift is given by their difference:
\begin{equation}
  \vec{\Delta}(\vec{K}_\perp) = \left(-\nabla_{\vec{k}_\perp} \phi
    +\left(\frac{\eta_-^2}{1+\eta_-^2}
    - \frac{\eta_+^2}{1+\eta_+^2}\right)\nabla_{\vec{k}_\perp} \alpha
    \right)_{\vec{k}_\perp = \vec{K}_\perp}.
\end{equation}
Using the equality $\eta_+ = 1/\eta_-$ yields Eq.~(6) of the main text.

\subsection{Non-Gaussian beams and gauge invariance}

In this section, we show that the same beam shift formulas can be derived even if the Gaussian envelope approximation is relaxed.  This derivation also shows that even though the reflection coefficient $\phi$ depends on the gauge choice used in the definition of the eigenfunctions [e.g., Eq.~(1) of the main text], the beam shift $\vec{\Delta}$ is a physical quantity that is gauge invariant.  The beam shifts are derived in terms of geometrical connections defined using the spinors of the incident and reflected beams.

Let $g(\vec{k}_\perp)$ be a real valued function peaked at $\vec{k}_\perp=\vec{K}_\perp$. The incident and reflected beam can be written as 
\begin{align}
  \Psi_i(\vec{r},\vec{K}_\perp) &= \int d\vec{k}_\perp \;
  g(\vec{k}_\perp-\vec{K}_\perp) \; \psi_i(\vec{k}_\perp)\;
  e^{i\vec{k}_\perp \cdot \vec{r}_\perp+ik_z^- z} \\
  \Psi_r(\vec{r},\vec{K}_\perp) &= \int d\vec{k}_\perp \;
  g(\vec{k}_\perp-\vec{K}_\perp) \; \mathrm{r}(\vec{k}_\perp)  \;
  \psi_r(\vec{k}_\perp)e^{i\vec{k}_\perp \cdot \vec{r}_\perp+ik_z^+ z}.
\end{align}
Here, $\mathrm r(\vec{k}_\perp)=e^{i\phi(\vec{k}_\perp)}$ is the reflection coefficient. To track the peak of the wavepacket, we calculate the probability amplitude at the interface $z=0$:
\begin{equation}
  \left|\Psi_{i(r)} (\vec{K}_\perp,\vec{r}_\perp)\right|^2
  = \int d\vec{k}_\perp d\vec{k}^\prime_\perp \; 
  g_F(\vec{k}_\perp,\vec{k}^\prime_\perp,\vec{K}_\perp) 
  \; e^{i\theta_{i(r)}(\vec{k}_\perp,\vec{k}^\prime_\perp)}
\end{equation}
where 
\begin{align}
  g_F(\vec{k}_\perp,\vec{k}^\prime_\perp,\vec{K}_\perp)
  &= g(\vec{k}_\perp,\vec{K}_\perp)g(\vec{k}^\prime_\perp,\vec{K}_\perp), \\
  \theta_i(\vec{k}_\perp,\vec{k}^\prime_\perp, \vec{r})
  &= -i \log \left\langle\psi_i(\vec{k}_\perp)\right| \left.\psi_i(\vec{k}^\prime_\perp)\right\rangle
  + (\vec{k}_\perp-\vec{k}^\prime_\perp) \cdot \vec{r}\\
  \theta_r(\vec{k}_\perp,\vec{k}^\prime_\perp, \vec{r})
  &= -i \log \left\langle\psi_r(\vec{k}_\perp) \right.\left|\psi_r(\vec{k}^\prime_\perp)\right\rangle
  + (\vec{k}_\perp-\vec{k}^\prime_\perp) \cdot \vec{r} 
  + \phi(\vec{k}_\perp)-\phi(\vec{k}^\prime_\perp).
\end{align}
We have used Dirac's bra-ket notation to express the various $k$-space integrals.  The peak of the probability amplitude in real space is the stationary point $\vec{R}$ determined by
\begin{equation}
\nabla_{\vec{k}_\perp} \theta_{i(r)}(\vec{k}_\perp,\vec{k}^\prime_\perp,\vec{R}_{i(r)} )\Big|_{\vec{K}_\perp} = 0
\end{equation}
This gives the peaks of the incident and reflected beams:
\begin{align}
  \vec{R}_i &= \vec{\cal A}_i(\vec{K}_\perp) \\
  \vec{R}_r &= \vec{\cal A}_r(\vec{K}_\perp) - \nabla_{\vec{k}_\perp} 
  \phi(\vec{k}_\perp)|_{\vec{K}_\perp}.
\end{align}
Here,
\begin{equation}
\vec{\cal A}_{i(r)}(\vec{k}_\perp)=i \Big\langle \psi_{i(r)}(\vec{k}_\perp) \Big| \nabla_{\vec{k}_\perp} \Big| \psi_{i(r)}(\vec{k}_\perp) \Big\rangle
\end{equation}
is the Berry connection for the incident (reflected) wave. By explicit calculation using the eigenstates of Weyl Hamiltonian, we obtain
\begin{align}
  \vec{\cal A}_{i}(\vec{k}_\perp)
  &=-\frac{\eta_-^2}{\eta_-^2+1}\nabla_{\vec{k}_\perp}\alpha \\
  \vec{\cal A}_{r}(\vec{k}_\perp)
  &= -\frac{\eta_+^2}{\eta_+^2+1}\nabla_{\vec{k}_\perp}\alpha.
\end{align}
The shift is then given by 
\begin{align}
  \vec{\Delta}(\vec{K}_\perp)
  &= \vec{R}_r-\vec{R}_i=\vec{\cal A}_r(\vec{K}_\perp)
  - \vec{\cal A}_i(\vec{K}_\perp)
  - \nabla_{\vec{k}_\perp} \phi(\vec{k}_\perp)|_{\vec{K}_\perp} \\ 
  &=\left(\frac{\eta_-^2-1}{\eta_-^2+1}\nabla_{\vec{k}_\perp} \alpha 
    - \nabla_{\vec{k}_\perp} \phi \right)_{\vec{K}_\perp}.
  \label{Shift_berry}
\end{align}
We have used the fact that $\eta_- = 1/\eta_+$.  Eq.~\eqref{Shift_berry} shows that Eq. 6 of the main text is valid for any shape of incident beam as long as the beam profile remains a real function peaked at a single point in real space. This in turn implies that the neglected higher order terms, which generally cause a change of the beam profile, will not change the calculated shift even upon multiple reflections.

Now consider boundary conditions of the form
\begin{equation}
  \mathrm{M}(\vec{k}_\perp)\big[\psi_i(\vec{k}_\perp)
    + r(\vec{k}_\perp)\psi_r(\vec{k}_\perp)\big] = 0,
\end{equation}
where $\mathrm M(\vec{k}_\perp)$ is a $2 \times 2$ matrix characterizing the boundary.  The reflection phase can be written as
\begin{equation}
\phi(\vec{k}_\perp)=i\log[\mathrm M\psi_r(\vec{k}_\perp)] - i \log [\mathrm M\psi_i(\vec{k}_\perp)].
\end{equation} 
The derivative is
\begin{equation}
  \nabla \phi = i\frac{\nabla \mathrm M\psi_r}{\mathrm M\psi_r}
  - i\frac{\nabla \mathrm M\psi_i}{\mathrm M\psi_i}.
  \label{gradphi_geom}
\end{equation}
We define two new objects
\begin{equation}
\psi_{i(r)}^{\theta} = \mathrm M\psi_{i(r)},
\end{equation}
which are normalized as $\big\langle\psi_{i(r)}^{\theta}(\vec{k}_\perp) \big| \psi_{i(r)}^{\theta}(\vec{k}_\perp)\big\rangle=1$.  Then Eq.~\eqref{gradphi_geom} reads
\begin{equation}
  \nabla \phi = \Big\langle \psi_{r}^{\theta}(\vec{k}) \Big| i\nabla_{\vec{k}} \Big| \psi_{r}^{\theta}(\vec{k})\Big\rangle
  - \Big\langle \psi_{i}^{\theta}(\vec{k}) \Big| i\nabla_{\vec{k}} \Big| \psi_{i}^{\theta}(\vec{k})\Big\rangle,
\end{equation}
which is the difference between two geometric (non-Berry) connections. These geometric connections are related to the parallel transport determined by the boundary condition.
 
Although both $\vec{R}_i$ and $\vec{R}_r$ are gauge dependent (which amounts to a different choice of origin for the individual waves), their difference is gauge invariant. When
\begin{equation}
  \psi_{i(r)}(\vec{k}_\perp) \rightarrow
  \psi_{i(r)}(\vec{k}_\perp) \exp({i\chi_{i(r)}(\vec{k}_\perp)}),
\end{equation}
we find that
\begin{align}
  \vec{\cal A}_{i(r)}(\vec{k}_\perp)
  &\rightarrow \vec{\cal A}_{i(r)} (\vec{k}_\perp)- \nabla_{\vec{k}_\perp} 
  \chi(\vec{k}_\perp) \\
  \nabla_{\vec{k}_\perp}\phi(\vec{k}_\perp)
  &\rightarrow \nabla_{\vec{k}_\perp}\phi(\vec{k}_\perp)
  - \nabla_{\vec{k}_\perp}\chi_{r}(\vec{k}_\perp)
  + \nabla_{\vec{k}_\perp} \chi_{i} (\vec{k}_\perp).
\end{align}
Hence, the shift given by Eq.~\eqref{Shift_berry} is gauge invariant.

The shift itself is given by
\begin{equation}
  \vec{\Delta}(\vec{K}_\perp)
  = \left( \frac{\eta_-^2-1}{\eta_-^2+1}\nabla_{\vec{k}_\perp}
  \alpha(\vec{k}_\perp)
  - \nabla_{\vec{k}_\perp} \phi(\vec{k}_\perp)\right)_{\vec{k}_\perp = \vec{K}_\perp}.
\end{equation}
The reflection coefficient is given by
\begin{equation}
  r = e^{i\phi} = -\frac{1+\eta_- e^{i\beta}}{\eta_- + e^{i \beta}},
\end{equation}
where $\beta(\vec{k}_\perp)=\alpha(\vec{k}_\perp)-\theta_b$. Therefore,
\begin{align}
  \nabla \phi &= -i \frac{\nabla r(\vec{k}_\perp)}{r(\vec{k}_\perp)} \\
  &= -\frac{i}{r} \frac{ie^{i\beta}(\eta_-^2-1)(\nabla \beta)
    + (e^{2i\beta}-1)(\nabla\eta_-)}{(\eta_-+e^{i\beta})^2} \\
  &= \frac{(\eta_-^2-1)(\nabla \beta)+2
    \sin \beta (\nabla \eta_-)}{1+\eta_-^2+2\eta_-\cos \beta},
    \label{Grad_phi}
\end{align}
and
\begin{align}
  \nabla\eta_-
  &= \frac{-E}{\eta (E+k_z^-)^2} \frac{\vec{k}_\perp}{|k_z^-(\vec{k}_\perp)|} \\
  \nabla \beta &= \nabla \alpha = (-k_y,k_x)/|\vec{k}_\perp|^2.
\end{align}
Putting everything together, we obtain the explicit formula
\begin{equation}
\vec{\Delta}(\vec{K}_\perp) = \left[\frac{\eta_-^2(\vec{K}_\perp)-1}{\eta_-^2(\vec{K}_\perp)+1} + \frac{k_z^-(\vec{K}_\perp)}{E+|\vec{K}_\perp| \cos \beta(\vec{K}_\perp)}\right]\frac{1}{|\vec{K}_\perp|^2} (-K_y,K_x) + \frac{E \sin \beta(\vec{K}_\perp)}{|\vec{K}_\perp| E + |\vec{K}_\perp|^2 \cos \beta(\vec{K}_\perp)}\frac{\vec{K}_\perp}{|k_z(\vec{K}_\perp)|}.
\label{Shift_analytical}
\end{equation}

\subsection{Inverse square root scaling of the beam displacement}

\begin{figure}
  {\centering \includegraphics[width=0.8\textwidth]{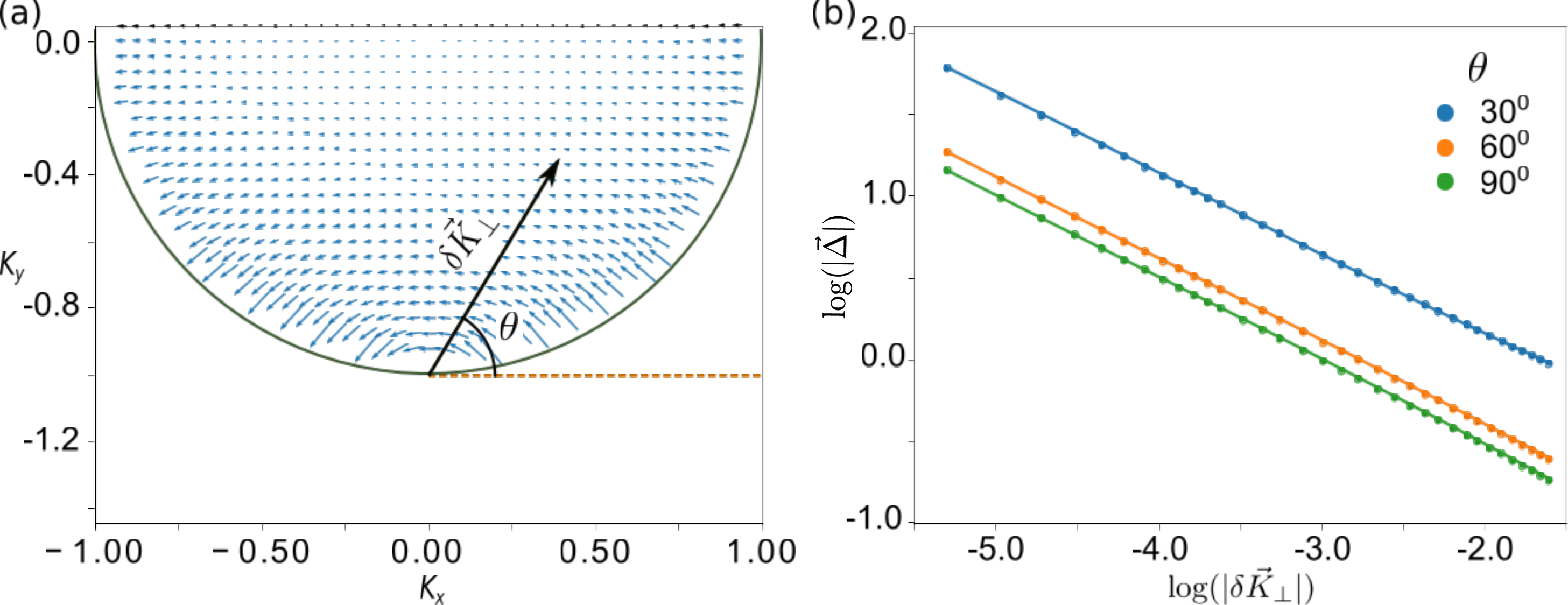}}
  \caption{(a) Zoomed-in image of the shift vector $\vec{\Delta}$ close to the Fermi arc touching-point.  (b) Log-log plot of the magnitude of the shift vector in different directions away from the Fermi arc touching-point.  The straight lines show numerical least squares fits, which indicate that the magnitudes indeed scale as $|\vec{\Delta}| \sim |\delta \vec{K}_\perp|^{-1/2}$. }
  \label{Scaling}
\end{figure}

\begin{figure}
  {\centering \includegraphics[width=0.8\textwidth]{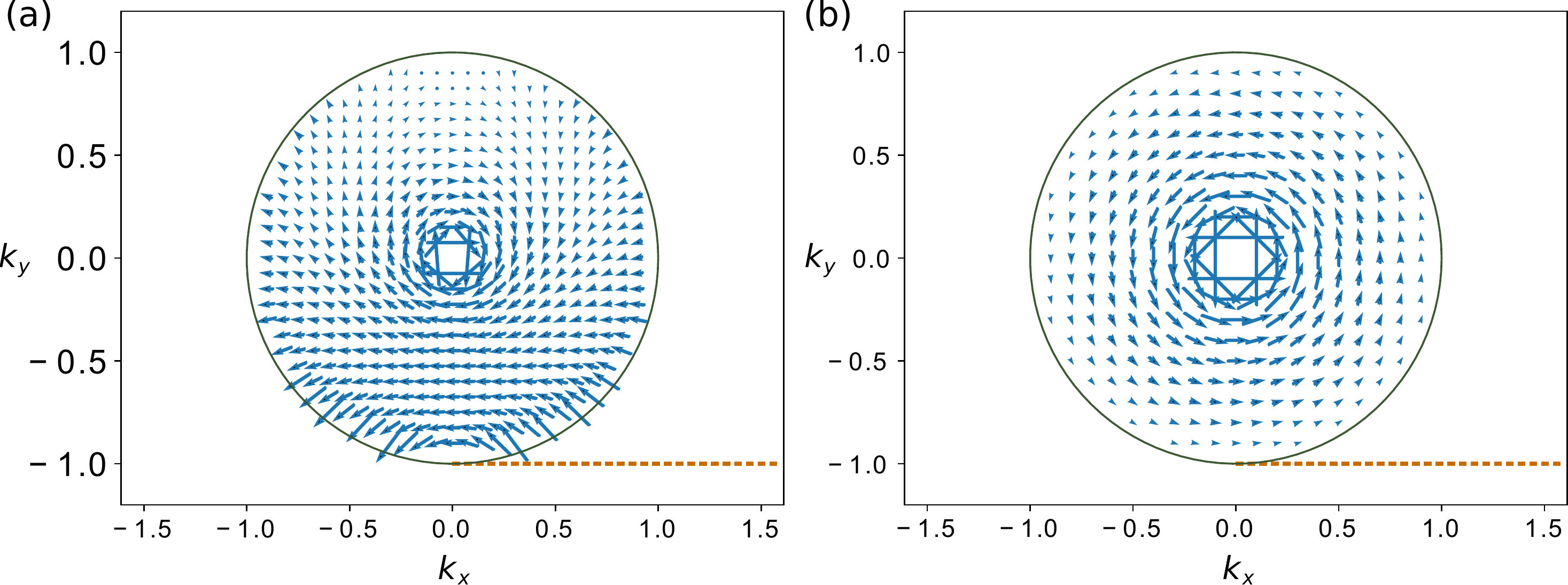}}
  \caption{Momentum dependence of the vector fields (a) $-\nabla \phi$ and (b) $\nabla \alpha$, for $\theta_b = \pi/2$. The half-vortex structure near the Fermi arc touching-point is evidently due to the gradient of $\phi$.  Note that $-\nabla \phi$ and $\nabla \alpha $ circulate in opposite directions around the origin.}
  \label{Grad}
\end{figure}

Fig.~\ref{Scaling} shows the inverse square-root scaling of the magnitude of the shift near the Fermi arc touching-point $\vec{K}_{\mathrm{fa}}$.
As mentioned in the main text, the vortex structure near the Fermi arc touching-point is due to the gradient of the reflection phase, as explicitly shown in Fig.~\ref{Grad}. The scaling can be verified by expanding the gradient of reflection phase about the Fermi arc touching-point.  The latter is derived by requiring the penetration constant of the Fermi arc surface state to vanish, which yields
\begin{equation}
\vec{K}_{\mathrm{fa}} = \Big(-E\cos(\theta_b),-E\sin(\theta_b)\Big).
\end{equation}
We define $\vec{k}_\perp = \vec{K}_{\mathrm{fa}} + \delta \vec{k}$, where $\delta \vec{k}=(\delta k_x, \delta k_y)$ is the wave-vector measured from the Fermi arc touching-point. The shift $\vec{\Delta}$, given by \eqref{Shift_analytical}, is to be expanded up to linear order in $\delta \vec{k}$.

We can expand $k_z^-$ in linear order as follows:
\begin{equation}
k_z^- \approx -\sqrt{2E[\delta k_x \cos(\theta_b) + \delta k_y 
      \sin(\theta_b)]} = - Eq 
\end{equation}
 where
\begin{equation}
  q =\sqrt{\frac{2[\delta k_x \cos(\theta_b) + \delta k_y \sin(\theta_b)]}{E}}.
\end{equation}
And from the definition of $\alpha$ we  have
\begin{align}
\begin{aligned}
  \cos(\alpha) \approx  \frac{p\sin(\theta_b)}{E} - \cos(\theta_b), \;\;
  \sin(\alpha) \approx -\frac{p\cos(\theta_b)}{E} - \sin(\theta_b),  
\end{aligned}
\end{align}
where
\begin{equation}
  p = \delta k_x \sin(\theta_b) - \delta k_y \cos (\theta_b).
\end{equation}
and this gives
\begin{align}
\begin{aligned}
\sin \beta \approx -\frac{p}{E}, \;\; \cos \beta \approx -1
\end{aligned}
\end{align}
where $\beta = \alpha - \theta_b$. Now, from \eqref{Grad_phi} and \eqref{Shift_analytical} we have
\begin{align}
\begin{aligned}
\nabla \phi \approx -\frac{-q}{(1-\sqrt{1-q^2})}\frac{1}{E^2(1-q^2)} (-\delta k_y+E\sin \theta_b, \delta k_x-E\cos \theta_b) \\ 
- \frac{p}{E^2(\sqrt{1-q^2}-(1-q^2))}\frac{1}{Eq} (\delta k_x - E\cos \theta_b, \delta k_y - E\sin \theta_b).
\end{aligned}
\end{align}
which expanding up to linear order in $q$ simplifies to 
\begin{equation}
\nabla \phi \approx -\frac{2}{E^2}(\frac{1}{q} +q) (-\delta k_y+E\sin \theta_b, \delta k_x-E\cos \theta_b) - \frac{2p}{E^3q^3}(\delta k_x - E\cos \theta_b, \delta k_y - E\sin \theta_b)
\end{equation}
Noting that $p/q^3 \sim 1/q$, it is readily seen that the leading order term scales as $1/q$.
%
%

\subsection{Derivation of the shift for the quadratic Hamiltonian}

The quadratic Hamiltonian
\begin{equation}
  H = v \begin{bmatrix}
    -k_y & \gamma(k_x^2-m)-i  k_z  \\
    \gamma(k_x^2-m)+i  k_z &  k_y  \\
  \end{bmatrix}
\end{equation}
has dispersion relation
\begin{equation}
  E = \pm v\sqrt{(k_x^2-m)^2 + k_y^2 + k_z^2}.
\end{equation}
This can exhibit either paired Weyl points or complete band-gap, depending on the choice of $m$.  In the region $z > 0$, we set $m = m_0 > 0$, so that there is a pair of Weyl points at $\vec{k} = [\pm\sqrt{m_0}, \,0, \, 0]$; for small values of $|E|$, there are two distinct Weyl cones which merge as $|E|$ increases.  In the region $z < 0$, we set $m = -m_1 < 0$ so that there is a complete band gap in the range $|E| < v\gamma |m_1|$.
In the Weyl medium, for incident and reflected plane waves, the eigenvectors reads
\begin{equation}
  \Psi_i = \frac{1}{\sqrt{1+\eta^2}}
  \begin{bmatrix}
    e^{-i\alpha_-}\\
    \eta 
  \end{bmatrix}, \;\;
  \Psi_r = \frac{e^{i\phi}}{\sqrt{1+\eta^2}}
  \begin{bmatrix}
    e^{-i\alpha_+}\\
    \eta
  \end{bmatrix},
  \label{eigirquad}
\end{equation}
where
\begin{align}
  \begin{aligned}
  \alpha_{\pm} &= \tan^{-1}\left(\frac{k_z^{\pm}}{\gamma (k_x^2-m_0)}\right), \;\;
  \eta = \sqrt{\frac{E+vk_y}{E-vk_y}}, \\
  k_z^{\pm} &= \pm \sqrt{(E/v)^2 - \gamma^2 (k_x^2-m_0)^2 -k_y^2},
  \end{aligned}
\end{align}
where as before $k_z < 0$ ($k_z > 0$) branch chosen for the incident (reflected) wave.

The reflection amplitude $e^{i\phi}$ is calculated by matching the total incident and reflected envelope functions at $z=0$ plane with the evanescent wave in the band-gap medium given by
\begin{equation} 
\psi_{t} = \frac{t}{N_t}\begin{bmatrix}
     E/v-k_y\\
     \gamma(k_x^2+m_1)+\kappa
   \end{bmatrix} e^{ik_{x}x+ik_{y} y +\kappa z},
   \label{IIWave}
\end{equation}
where $\kappa=\sqrt{\gamma^2(k_x^2+m_1)^2+k_y^2-(E/v)^2}$ is the inverse decay length of the evanescent field, $t$ is the transmission coefficient and $N_t$ is a normalization factor.
Equating $\psi_{i} + \psi_{r} = \psi_{t} $ at $z=0$ for all $x$ and $y$, the reflection coefficient $e^{i\phi}$ is calculated:
\begin{equation} 
e^{i\phi} = \frac{\eta(E-v k_y)-[v\gamma(k_x^2+m_1)+v\kappa]e^{-i\alpha_-}}{e^{-i\alpha_+}[v\gamma(k_x^2+m_1)+v \kappa]-\eta(E-vk_y)}
\label{refamp_quad}
\end{equation}  
The spatial shift in the reflected beam can be calculated in a manner similar to the previous case, which gives
\begin{equation}
  \vec{\Delta}(\vec{K}_\perp) = \left(-\nabla_{\vec{k}_\perp} \phi 
  -\frac{2}{\eta^2+1}\nabla_{\vec{k}_\perp} \alpha_-
  \right)_{\vec{k}_\perp = \vec{K}_\perp}.
  \label{shift_quad} 
\end{equation}
The results are shown in Fig.~3 of the main text.

\subsection{Consecutive reflections in a thin film geometry}

\begin{figure}
  {\centering \includegraphics[width=1\textwidth]{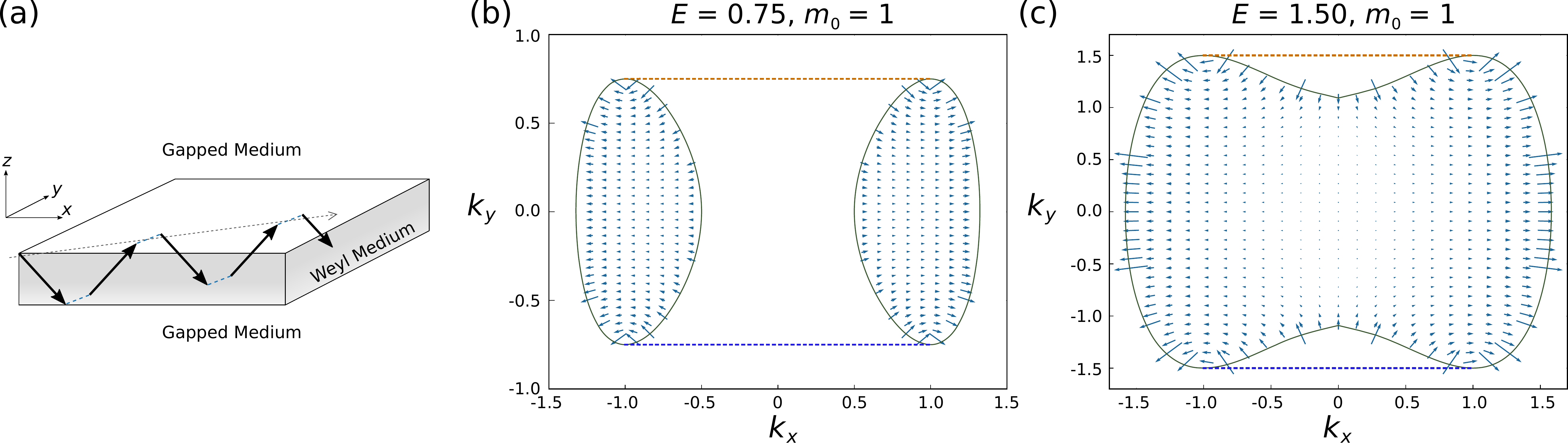}}
  \caption{(a) Schematic of a beam bouncing within a thin film geometry. (b)--(c) Beam shift versus incident in-plane momentum, over two consecutive reflections in the thin film.  The orange (blue) dashes show the Fermi arc on the bottom (top) surface.}
  \label{Bounce_film}
\end{figure}

Consider a film of Weyl medium of thickness $2L$, within the space $|z| < L$ bounded above and below by a gapped medium.  In this geometry, a beam in the Weyl medium will reflect repeatedly off the two parallel surfaces, much like a bouncing-ball trajectory within a waveguide, as shown in Fig.~\ref{Bounce_film}(a).

As the beam reflects consecutively off the bottom and top surfaces of the film, the beams displacements accumulate instead of cancelling.  Fig.~\ref{Bounce_film}(b)--(c) plots the numerically-calculated beam shift over two consecutive reflections, using the quadratic Weyl Hamiltonian with continuity boundary conditions at $|z| = L$.  The shift from the upper surface can be found by replacing $k_z^{\pm} \rightarrow k_z^{\mp}$ and $\kappa \rightarrow -\kappa$ in the previous equations.

In explicit terms, we take the Hamiltonian
\begin{equation}
H = 
\begin{bmatrix}
- k_y & (k_x^2 - m) - i k_z \\
(k_x^2 - m) + i k_z & k_y
\end{bmatrix},
\quad
m=
\begin{cases}
m_0 > 0 , & |z| \leq L\; ({\textrm{Weyl~medium}})  \\
- m_1 ,  & |z| > L \; ({\textrm{gapped~medium}}).
\end{cases}
\end{equation}
Inside the Weyl semimetal, we look for states $\propto e^{i \vec{k}_\perp \cdot \vec{r}_\perp + i k_z^\pm z } $ with $\vec{k}_\perp = (k_x, k_y)$ and $\vec{r}_\perp = (x,y)$. We focus on a fixed energy $E$, whose bulk states read
\begin{align}
\begin{aligned}
  \psi_{\pm}^{\mathrm{w}} (\vec{k}_\perp, \vec{r}_\perp, z)
  &= \frac{1}{\sqrt{1+\eta^2}}
  \begin{bmatrix}
    e^{i \alpha_\mp} \\ \eta 
  \end{bmatrix} 
  e^{i \vec{k}_\perp \cdot \vec{r}_\perp  + i k_z^\pm z } \\
  \eta &= \sqrt{\frac{E + k_y }{E - k_y}} \\
  \alpha_\pm &= \arctan \Big( \frac{k_z^\pm}{k_x^2 -m_0} \Big) \\
  k_z^\pm &= \pm \sqrt{E^2 - (k_x^2 -m_0)^2 - k_y^2}.
\end{aligned}
\end{align}
Here, $\psi_\pm^{\mathrm{w}} (\vec{k}_\perp, \vec{r}_\perp, z)$ propagates to the upper/lower surface $z = \pm L$.  We assume that the energy lies within the gap of the external medium.  In the $z  > L$ ($z < -L)$ region, the solution has the decaying form
\begin{equation}
  \psi_\pm^{\mathrm{evs}} (\vec{k}_\perp, \vec{r}_\perp, z) \propto
  \begin{bmatrix}
    E - k_y \\ k_x^2 + m_1 \mp  \kappa  
  \end{bmatrix} 
  e^{i k_x x + i k_y \mp \kappa z }.
\end{equation}
This wave function is not normalized, and due to energy conservation,
\begin{equation}
\kappa = \sqrt{ (k_x^2 + m_1)^2 - (k_x^2 - m_0)^2 }.
\end{equation}
At the interface $z=\pm L$, the incident wave $\psi_\pm^{\mathrm{w}}$, the reflected wave  $\psi_\mp^{\mathrm{w}}$, and the evanescent wave $\psi_\pm^{\mathrm{evs}}$ must match:
\begin{equation}
  \psi_\pm^{\mathrm{w}} (\vec{k}_\perp, \vec{r}_\perp, \pm L) + \mathrm{r}_{\pm} (\vec{k}_\perp) \psi_\mp^{\mathrm{w}} (\vec{k}_\perp, \vec{r}_\perp, \pm L) =\mathrm{t}_\pm (\vec{k}_\perp) \psi_\pm^{\mathrm{evs}} (\vec{k}_\perp, \vec{r}_\perp, \pm L),
\end{equation}
where $\mathrm{r}_{\pm} (\vec{k}_\perp)$ and $\mathrm{t}_{\pm} (\vec{k}_\perp)$ are reflection and transmission coefficients at $z = \pm L$.  By multiplying $(k_x^2 + m_1 \mp  \kappa, - E + k_y )$ to both sides of the equation, we obtain the total reflection coefficient
\begin{equation}
\mathrm{r}_\pm (\vec k)  = -
\frac{e^{ i \alpha_\mp } (k_x^2+ m_1 \mp \kappa) -  
\sqrt{ E^2 - k_y^2 }   }
{e^{i \alpha_\pm } (k_x^2+ m_1 \mp \kappa) - \sqrt{E^2 - k_y^2 } }
\equiv - \exp [i \phi_\pm (\vec{k}_\perp)].
\end{equation}
For wave packets composed by $\psi_\pm^{\mathrm{w}}$, the shift between the total reflected wave and the incident wave in real space at the interface $z = \pm L$ is
\begin{equation}
\vec{\Delta}_\pm (\vec{k}_\perp ) =  \vec A_\mp (\vec{k}_\perp ) - \vec A_\pm (\vec{k}_\perp ) - \nabla_{\vec{k}_\perp} \phi_\pm (\vec{k}_\perp)
\end{equation}
where
$\vec{A}_\pm (\vec{k}_\perp) = i \big\langle \psi_{\pm}^{\mathrm{w}}(\vec{k}_\perp) \big| \nabla_{\vec{k}_\perp} \big| \psi_{\pm}^{\mathrm{w}}(\vec{k}_\perp)\big\rangle$ is the Berry connection for the incident/reflected state.  Since $\vec{A}_+ (\vec{k}_\perp) - \vec{A}_- (\vec{k}_\perp)$ and $\vec{A}_-(\vec{k}_\perp) - \vec{A}_+(\vec{k}_\perp)$ cancel, after two consecutive reflections between opposite surfaces $z = \pm L$, the total shift is
\begin{equation}
\vec{\Delta}_{\mathrm{tot}} (\vec{k}_\perp ) = 
  \vec{\Delta}_+ (\vec{k}_\perp ) + \vec{\Delta}_- (\vec{k}_\perp )
  = - \nabla_{\vec{k}_\perp} \phi_+ (\vec{k}_\perp) -\nabla_{\vec{k}_\perp} \phi_- (\vec{k}_\perp).
\end{equation}
The resulting map of the shift is shown in Fig.~\ref{Bounce_film}.

\end{widetext}

\end{document}